\documentclass[preprint,superscriptaddress,eqsecnum,nofootinbib]{revtex4}
\usepackage{graphicx,amssymb,amsbsy,bm,amsmath,psfrag,dsfont}
\usepackage{hyperref}

\begin{document}

\newcommand{\be}{\begin{equation}}
\newcommand{\ee}{\end{equation}}
\newcommand{\bea}{\begin{eqnarray}}
\newcommand{\eea}{\end{eqnarray}}
\newcommand{\no}{\noindent}

\newcommand{\la}{\lambda}
\newcommand{\si}{\sigma}
\newcommand{\vp}{\mathbf{p}}
\newcommand{\vk}{\vec{k}}
\newcommand{\vx}{\vec{x}}
\newcommand{\om}{\omega}
\newcommand{\Om}{\Omega}
\newcommand{\ga}{\gamma}
\newcommand{\Ga}{\Gamma}
\newcommand{\gaa}{\Gamma_a}
\newcommand{\al}{\alpha}
\newcommand{\ep}{\epsilon}
\newcommand{\app}{\approx}
\newcommand{\uvk}{\widehat{\bf{k}}}
\newcommand{\OM}{\overline{M}}

\title{On Neutrino Flavor States}
\author{Chiu Man Ho} \email{chiuman.ho@vanderbilt.edu}
  \affiliation{Department of
  Physics and Astronomy, Vanderbilt University, Nashville, Tennessee
  37235, USA}
\date{\today}

\begin{abstract}
We review the issues associated with the construction of neutrino flavor states. We then provide a consistent proof that the flavor
states are approximately well-defined only if neutrinos are ultra-relativistic or the mass differences are negligible compared to energy.
However, we show that weak interactions can be consistently described by only neutrino mass eigenstates. Meanwhile, the second quantization of
neutrino flavor fields generally has no physical relevance as their masses are indefinite. Therefore, the flavor states are not physical
quantum states and they should simply be interpreted as definitions to denote specific linear combinations of mass eigenstates involved in
weak interactions. We also briefly discuss the implication of this work for the mixing between active and heavy sterile neutrinos.
\end{abstract}
\maketitle

\section{Introduction}
\label{sec:intro}

In the standard treatment of neutrino oscillation in the vacuum, the neutrino fields in the flavor and mass bases
are related by a unitary transformation
\bea
\label{unitary}
\nu_\al =\sum_{k}\, \mathcal{U}_{\al k}\, \nu_k\,,
\eea
where $\al=e,\mu,\tau$ and $k=1,2,3$ are the indices for the flavor and mass fields respectively. For the usual three neutrino case,
the unitary transformation is described by the Maki-Nakagawa-Sakata-Pontecorvo (MNSP) matrix:
\bea
\mathds{U} \equiv \left(
      \begin{array}{ccc}
        \mathcal{U}_{e 1} & \mathcal{U}_{e 2} & \mathcal{U}_{e 3} \\
        \mathcal{U}_{\mu 1} & \mathcal{U}_{\mu 2} & \mathcal{U}_{\mu 3} \\
        \mathcal{U}_{\tau 1} & \mathcal{U}_{\tau 2} & \mathcal{U}_{\tau 3} \\
      \end{array}
    \right) =\left(
               \begin{array}{ccc}
                 c_{21}\,c_{13} & s_{12}\,c_{13} & s_{13}\,e^{-i\,\delta} \\
                 -s_{21}\,c_{32} -c_{21}\,s_{32}\,s_{13}\,e^{i\,\delta} & c_{21}\,c_{32}-s_{21}\,s_{32}\,s_{13}\,e^{i\,\delta} & s_{32}\,c_{13} \\
                 s_{21}\,s_{32} -c_{21}\,c_{32}\,s_{13}\,e^{i\,\delta} & -c_{21}\,s_{32} -s_{21}\,c_{32}\,s_{13}\,e^{i\,\delta} & c_{32}\,c_{13} \\
               \end{array}
             \right)
\eea
where $c_{jk}=\cos(\theta_{jk})$, $s_{jk}=\sin(\theta_{jk})$ and $\delta$ is the CP-violating phase. The free Dirac equation for the neutrino flavor fields is given by
\bea
\label{FlavorDirac}
\left(\,i\,\!\!\not\!\partial - M_{\al\beta}\,\right) \, \nu_\beta =0\,,
\eea
where $\mathds{M}$ is the $3\times 3$ neutrino mass mixing matrix. Upon the unitary transformation in Eq. \eqref{unitary}, this Dirac equation is diagonalized to become
\bea
\label{MassDirac}
\left(\,i\,\!\!\not\!\partial - m_k \,\delta_{kj}\,\right) \, \nu_j =0\,,
\eea
with $m_k \,\delta_{kj} = \mathcal{U}^{\ast}_{\al k}\, M_{\al \beta}\,\mathcal{U}_{\beta j}$.

In general, the neutrino flavor state is given by
\bea
\label{FlavorState}
|\nu_\al \rangle =\sum_{k}\, \mathcal{U}_{\al k}^\ast\, |\nu_k \rangle \,.
\eea
At any time $t$, the flavor state $|\nu_\al \rangle $ evolves into
$|\nu_\al (t) \rangle =\sum_{k}\, e^{-i\,E_k\,t}\,\mathcal{U}_{\al k}^\ast\, |\nu_k \rangle$.
Then, the transition probability from the flavor
state $|\nu_\al (t)\rangle$ to the flavor state $|\nu_\beta \rangle$ is obtained by calculating
$P_{\al\beta}= |\langle \nu_\beta|\nu_\al(t) \rangle|^2$.
But the question is: How to derive Eq. \eqref{FlavorState} from the first principles of quantum field theory?

One could try to answer this question by starting with the second quantization of the neutrino fields. In the mass basis, the neutrino
field is given by

\bea
\label{massfield}
\nu_k(x) = \int\,\frac{d^3\,\mathbf{p}}{(2\pi)^{3/2}}\,\frac{1}{\sqrt{2\,E_k}}\,\sum_{s}\, \left(\,a_k(\mathbf{p},s)\,u_k(p,s)
\,e^{-i\,p\,x} +b^{\dagger}_k(\mathbf{p},s)\,v_k(p,s)\,e^{i\,p\,x}\,\right)\,,
\eea
where $E_k=\sqrt{\vp^2+m_k^2}$ is the energy of the neutrino mass fields. The second quantization of the neutrino mass fields $\nu_k$ according to Eq. \eqref{massfield} is
consistent with the Dirac equation in Eq. \eqref{MassDirac}, and so $m_k$ is a physically definite quantity.
The creation and annihilation operators obey the usual anticommutation relations:
\bea
\label{commute}
\{a_k(\mathbf{p},s),\, a_k^{\dagger}(\mathbf{q},s')\} = \{b_k(\mathbf{p},s),\, b_k^{\dagger}(\mathbf{q},s')\} = \delta^{(3)}(\mathbf{p}-\mathbf{q})\,\delta_{ss'}\,.
\eea
The spinors $u_k(p,s)$ and $v_k(p,s)$ are respectively the positive and negative energy solutions of the Dirac equation with
mass $m_k$. They are given by \cite{Peskin}
\bea
\label{spinors}
u_k(p,s)= \left(
              \begin{array}{c}
                \sqrt{p\cdot \sigma}\,\xi_s \\
                \sqrt{p\cdot \bar{\sigma}}\,\xi_s \\
              \end{array}
            \right)~~~~;~~~~
v_k(p,s)= \left(
              \begin{array}{c}
                \sqrt{p\cdot \sigma}\,\eta_s \\
                -\sqrt{p\cdot \bar{\sigma}}\,\eta_s \\
              \end{array}
            \right)\,,
\eea
where
\bea
\sigma^\mu= (1, \vec{\sigma})~~~~;~~~~\bar{\sigma}^\mu= (1, -\vec{\sigma}) ~~~~;~~~~
\xi_1=\eta_1= \left(
                     \begin{array}{c}
                         1 \\
                         0 \\
                       \end{array}
                     \right)~~~~;~~~~
\xi_2=\eta_2= \left(
                     \begin{array}{c}
                         0 \\
                         1 \\
                       \end{array}
                     \right) \,.
\eea
It is understood that we take the positive root of each eigenvalues when taking
the square-root of a matrix. These spinors obey the following orthogonality conditions:
\bea
\label{ortho}
u_k^\dagger (p ,s)\,u_k(p,s^{\prime})
&=& v_k^\dagger(p,s)\,v_k(p,s^{\prime}) =2\, E_k\,\delta_{s, s^{\prime}} \,,\\
\bar{u}_k (p ,s)\,u_k(p,s^{\prime})
&=& -\,\bar{v}_k(p,s)\,v_k(p,s^{\prime}) = 2\,m_k\,\delta_{s, s^{\prime}} \,,\\
\bar{u}_k (p ,s)\,v_k(p,s^{\prime})
&=& \bar{v}_k(p,s)\,u_k(p,s^{\prime}) = 0 \,, \\
u_k^\dagger(p,s)\, v_k(-p,s^{\prime}) &=& v_k^\dagger(-p,s)\, u_k(p,s^{\prime})= 0\,,
\eea
where the notation $v_k(-p,s)$ means $v_k(E_k, -\vp,s)$. They also satisfy the completeness condition:
\bea
\label{complete}
\sum_s\, \left\{\, u_k(p ,s)\,u_k^\dagger(p,s)+ v_k(-p ,s)\,v_k^\dagger(-p,s)  \,\right\} = 2\, E_k\,,
\eea
which has been written in a form that is more convenient for later discussions.

One might expect that we could second-quantize the neutrino flavor fields in the similar way:
\bea
\label{flavorfield}
\nu_\al(x) \stackrel{?}{=} \int\,\frac{d^3\,\mathbf{p}}{(2\pi)^{3/2}}\,\frac{1}{\sqrt{2\,E_\al}}\,\sum_{s}\, \left(\,a_\al(\mathbf{p},s)\,u_\al(p,s)
\,e^{-i\,p\,x} +b^{\dagger}_\al(\mathbf{p},s)\,v_\al(p,s)\,e^{i\,p\,x}\,\right)\,,
\eea
where $E_\al \stackrel{?}{=} \sqrt{\vp^2+m_{\nu_\al}^2}$ could be the energy of the neutrino flavor fields. The annihilation operators,
the creation operators and the spinors associated with the neutrino flavor fields could satisfy analogous conditions in
Eqs. (\ref{commute}--\ref{complete}) with $m_k$ replaced by $m_{\nu_\al}$ wherever applicable.

Without asking what is $m_{\nu_\al}$
for a moment, one could use Eq. \eqref{unitary} to relate $\nu_\al$ and $\nu_k$. It is then tempting to assume that
\bea
\label{wrong}
\frac{1}{\sqrt{2\,E_\al}}\,a_\al(\mathbf{p},s)\,u_\al(p,s)\, e^{-i\,E_\al\,t}= \sum_k \,\frac{1}{\sqrt{2\,E_k}}\,\mathcal{U}_{\al k}\,a_k(\mathbf{p},s)\,u_k(p,s)\, e^{-i\,E_k\,t}\,.
\eea
This assumption was made by \cite{Giunti1992}, although they also assumed that $u_\al(p,s) = u_k(p,s)$ which
is unreasonable. The mistake of assuming $u_\al(p,s) = u_k(p,s)$ was corrected by Giunti in \cite{Giunti2003}. Anyway, the main point
in \cite{Giunti1992} was that Eq. \eqref{wrong} implies that
\bea
\{a_\al(\mathbf{p},s),\, a_\beta^{\dagger}(\mathbf{q},s')\} &=& \delta^{(3)}(\mathbf{p}-\mathbf{q})\,\delta_{ss'}\,
\frac{1}{\sqrt{2\,E_\al}}\,\frac{1}{\sqrt{2\,E_\beta}}\,\,e^{i\,(E_\al-E_\beta)\,t}
\nonumber \\
&& u_\al^\dagger(p,s)\,\left(\, \sum_k\,\frac{1}{2\,E_k}\,\,\mathcal{U}_{\al k}\,\mathcal{U}_{\beta k}^{\ast} \,u_k(p,s)\,u_k^\dagger(p,s') \,\right) \,u_\beta(p,s')\,,
\eea
which is obviously not equivalent to the required anticommutation relation $\{a_\al(\mathbf{p},s),\, a_\al^{\dagger}(\mathbf{q},s')\} = \delta^{(3)}(\mathbf{p}-\mathbf{q})\,\delta_{ss'}$. This means that $a_\al(\mathbf{p},s)$ and $a_\al^{\dagger}(\mathbf{q},s')$ lose the meanings
of annihilation and creation operators. One can easily verify that the same arguments apply to the anticommutator $\{b_\al(\mathbf{p},s),\, b_\al^{\dagger}(\mathbf{q},s')\}$ if one makes a similar assumption as in Eq. \eqref{wrong} for $b_\al(\mathbf{p},s)$. Thus, it was concluded in \cite{Giunti1992} that the Fock space of neutrino flavor states, which is generated by successive applications of the creation operators on the vacuum, does not exist.

Years later, Giunti pointed out that the assumption in Eq. \eqref{wrong} made by \cite{Giunti1992} was unjustified \cite{Giunti2003}.
The correct expression should be
\bea
\label{correct}
&&\frac{1}{\sqrt{2\,E_\al}}\,a_\al(\mathbf{p},s)\,u_\al(p,s)\, e^{-i\,E_\al\,t} \nonumber \\
&=& \sum_k \,\frac{1}{\sqrt{2\,E_k}}\,
\mathcal{U}_{\al k} \left(\,a_k(\mathbf{p},s)\,u_k(p,s)\, e^{-i\,E_k\,t}+b_k^\dagger(-\mathbf{p},s)\,v_k(-p,s)\, e^{i\,E_k\,t}\,\right)\,.
\eea
Using the orthogonality condition $u_\al^\dagger (p ,s)\,u_\al(p,s^{\prime})
=2\, E_\al\,\delta_{s, s^{\prime}}$, the anticommutation relations in Eq. \eqref{commute}, the unitarity relation $\sum_k\, \mathcal{U}_{\al k}\,\mathcal{U}^{\ast}_{\beta k} =\delta_{\al \beta}$ and the completeness condition in Eq. \eqref{complete}, one can easily verify that Eq. \eqref{correct} indeed leads to the required anticommutation relation $\{a_\al(\mathbf{p},s),\, a_\al^{\dagger}(\mathbf{q},s')\} = \delta^{(3)}(\mathbf{p}-\mathbf{q})\,\delta_{ss'}$. With similar corrections to $b_\al(\mathbf{p},s)$, one can also obtain $\{b_\al(\mathbf{p},s),\, b_\al^{\dagger}(\mathbf{q},s')\} = \delta^{(3)}(\mathbf{p}-\mathbf{q})\,\delta_{ss'}$.
As a result, $a^\dagger_\al(\mathbf{p},s)$ and $b^\dagger_\al(\mathbf{p},s)$ can now be interpreted as the one-particle and one-antiparticle
creation operators respectively. Giunti thus concluded that the arguments in \cite{Giunti1992} against the Fock space of neutrino flavor states
are inconsistent \cite{Giunti2003}.

Meanwhile, Blasone and Vitiello (BV) have attempted to construct a Fock space for neutrino flavor states \cite{BV1}. This work admitted that
there is a unique vacuum $|0 \rangle$ that is annihilated by $a_k(\vp,s)$ and $b_k(\vp,s)$. They second-quantized the neutrino flavor fields according to Eq. \eqref{flavorfield}. Later they realized that taking $|0 \rangle$ as the unique vacuum leads to inconsistencies within their framework \cite{BV2}. They thus gave up $|0 \rangle$ and started defining $|0_{\al} \rangle$ as the neutrino flavor vacuum. This scheme appears mathematically consistent.

However, in the above discussions, we have postponed an important question: What is $m_{\nu_\al}$? In order to second-quantize $\nu_\al$ according to Eq. \eqref{flavorfield}, it is implicitly assumed that the Dirac equation $\left(\,i\,\!\!\not\!\partial - m_{\nu_\al}\,\right) \, \nu_\al =0$ exists. But in reality, this Dirac equation with mass $m_{\nu_\al}$ may not have any physical relevance. Generally speaking, it is not even related to
$\left(\,i\,\!\!\not\!\partial - M_{\al\beta}\,\right) \, \nu_\beta =0$ in Eq. \eqref{FlavorDirac}. Of course, a trivial special case occurs when the off-diagonal elements $|M_{\al\beta}|$ is much smaller than the diagonal elements $|M_{\al\al}|$, which is essentially the case with negligible neutrino mixing. In this case, one can identify $m_{\nu_\al} \approx |M_{\al\al}|$, and the Dirac equation $\left(\,i\,\!\!\not\!\partial - m_{\nu_\al}\,\right) \, \nu_\al =0$ is approximately equivalent to Eq. \eqref{FlavorDirac}. Since $|M_{\al\beta}| \ll |M_{\al\al}|$, the mass mixing matrix is almost diagonal. This means that we have $m_{\nu_e} \approx |M_{ee}| \approx m_1$, $m_{\nu_\mu} \approx |M_{\mu\mu}| \approx m_2$ and $m_{\nu_\tau} \approx |M_{\tau\tau}| \approx m_3$. Hence, the mass $m_{\nu_\al}$ becomes approximately definite for the negligible neutrino mixing case. However, in general, $m_{\nu_\al}$ is actually an indefinite unphysical quantity in the sense that it does not appear in the original Lagrangian
describing neutrino physics.

It was then shown by \cite{FHY} that the indefiniteness of $m_{\nu_\al}$ implies that the neutrino flavor vacuum defined in \cite{BV2} is also indefinite, and so there could be an infinite set of Fock spaces of neutrino flavor states.
The authors in \cite{FHY} thus emphasized that if one were to accept the scheme in \cite{BV2}, the indefinite quantity $m_{\nu_\al}$ should be absent from any observed quantities. But it was shown by Giunti that if these neutrino flavor states were to play roles in weak interactions, the indefinite mass $m_{\nu_\al}$ will become an observable quantity \cite{Giunti2003}. Therefore, Giunti conclude that ``the Fock spaces of flavor neutrinos are ingenuous mathematical constructs without physical relevance" \cite{Giunti2003}.

In fact, apart from the indefiniteness of $m_{\nu_\al}$, there is another issue that plagues the scheme in \cite{BV2}. The problem is that
the neutrino flavor vacuum defined in \cite{BV2} is time-dependent and hence Lorentz invariance is manifestly broken. Recently,
BV and collaborators attempted to tackle this issue by proposing neutrino mixing as a consequence of neutrino interactions with an external
non-abelian gauge field \cite{BV_nonabelian}. Under this framework, the Lorentz violation of the neutrino flavor vacuum can be attributed to the presence of a fixed external field which specifies a preferred direction in spacetime. However, at the moment, there is not a single sign of such a non-abelian gauge field in neutrino experiments. They proposed that this scheme can be tested in the tritium decay, but again the indefinite mass $m_{\nu_\al}$ becomes an observable quantity. Also, given the current stringent bounds on Lorentz violations \cite{Kostelecky}, it is unclear whether this scheme will survive.

In summary, it is currently unclear how to construct a consistent and physically relevant Fock space of neutrino flavor states. But one could still ask whether we can at least define it approximately under some special cases. This is precisely our task in the next section.

\section{Approximate Neutrino Flavor States}

In this section, we would like to ask the following question: If one insists on second-quantizing the neutrino flavor fields (which have indefinite masses) according to Eq. \eqref{flavorfield} and thereby constructing the flavor states, then under what special cases will they be approximately well-defined? A similar consideration was initiated by \cite{Giunti1992}. They proved that neutrino flavor states are
approximately well-defined if neutrinos are ultra-relativistic ($m_k \ll |\vp|$) or degenerate ($|m_k^2-m_j^2| \ll |\vp|^2$).
However, due to the wrong assumptions of $u_\al(p,s) = u_k(p,s)$ and $a_\al(\vp,s) \propto \sum_k\, a_k(\vp,s)$ as in Eq. \eqref{wrong},
it is unclear whether their proof is still valid. The purpose of this section is to provide a revised proof which is based on the
correct treatment.

Since we are not interested in the time evolution of the neutrino states for the following arguments, it is sufficient to work in the
Schrodinger picture (which simplifies the expressions). In the momentum space, the Hermitian conjugate of the neutrino field in the
mass and flavor bases are respectively given by
\bea
\nu^\dagger_{k}(\vp) &=& \frac{1}{\sqrt{2\,E_k}}\,\sum_{s}\,\left\{\,a^\dagger_k(\vp,s)\,u^\dagger_k(p,s)+b_k(-\vp,s)\,v^\dagger_k(-p,s)\,\right\}\,, \\
\nu^\dagger_{\al}(\vp) &=& \frac{1}{\sqrt{2\,E_\al}}\,\sum_{s}\,\left\{\,a^\dagger_\al(\vp,s)\,u^\dagger_\al(p,s)+b_\al(-\vp,s)\,v^\dagger_\al(-p,s)\,\right\}\,.
\eea
Using the orthogonality conditions $u_\al^\dagger (p ,s)\,u_\al(p,s^{\prime})
=2\, E_\al\,\delta_{s, s^{\prime}}$ and $v_\al^\dagger(-p,s)\, u_\al(p,s^{\prime})= 0$, one can easily deduce that $\sqrt{2 \, E_\al}\, a^\dagger_\al(\vp,s) = \nu^\dagger_{\al}(\vp) \,u_\al(p,s) $.

Physically, there is only one vacuum in the real world. This vacuum should be an eigenstate of the free Hamiltonian for the neutrinos and,
at the same time, carry the lowest energy. But this is precisely the vacuum $|0 \rangle$ that is annihilated by $a_k(\vp,s)$ and $b_k(\vp,s)$.
Thus, the one-particle neutrino flavor state is given by $|\nu_\al(\vp,s)\,\rangle =\sqrt{2 \, E_\al}\, a^\dagger_\al(\vp,s) \, |0 \rangle =
\nu^\dagger_{\al}(\vp) \,u_\al(p,s)\,|0 \rangle $. Using Eq. \eqref{unitary}, we obtain
\bea
\label{FlavorGeneral}
|\nu_\al(\vp,s)\,\rangle = \sum_{k}\, \frac{1}{2\,E_k}\; \mathcal{U}^\ast_{\al k}\, \left(\, u^\dagger_k(p,s)\, u_\al(p,s) \,\right)\,|\nu_k(\vp,s)\,\rangle \,,
\eea
where $|\nu_k(\vp,s)\,\rangle =\sqrt{2 \, E_k}\, a^\dagger_k(\vp,s) \, |0 \rangle$. In general, there is no orthogonality condition between $u^\dagger_k(p,s)$ and $u_\al(p,s)$. So the above expression is generally not equivalent to the neutrino flavor state given in Eq. \eqref{FlavorState}. However, we will show that Eq. \eqref{FlavorGeneral} reduces to Eq. \eqref{FlavorState} in two special cases both of which approximately lead to $u^\dagger_k(p,s)\, u_\al(p,s) \approx 2\,E_k$:

\begin{itemize}
  \item Ultra-relativistic neutrinos $(\,m_k,\, m_{\nu_\al} \ll |\vp|\,)$:\\
  In this limit, we have $E_k \approx |\vp|$ and $E_\al \approx |\vp|$ which imply that
  \bea
  u^\dagger_k(p,s)\, u_\al(p,s) \approx \left(\,|\vp| - \vp\cdot \vec{\sigma} \right) +\left(\,|\vp| +\vp\cdot \vec{\sigma} \right) =2\,|\vp|\,.
  \eea
  As a result, Eq. \eqref{FlavorGeneral} reduces to $|\nu_\al(\vp,s)\,\rangle \approx \sum_{k}\, \mathcal{U}^\ast_{\al k}\,|\nu_k(\vp,s)\,\rangle$
  which is the neutrino flavor state given by Eq. \eqref{FlavorState}.

  \item Negligible mass differences compared to energy $(\,|m_k^2-m_{\nu_\al}^2| \ll E_k^2\,)$: \\
  In this limit, $E_\al = E_k\,\sqrt{1 + (\,m_{\nu_\al}^2-m_k^2\,)/E^2_k} \approx E_k$ which implies that
  \bea
  u^\dagger_k(p,s)\, u_\al(p,s) \approx \left(\,E_k - \vp\cdot \vec{\sigma} \right) +\left(\,E_k +\vp\cdot \vec{\sigma} \right) =2\,E_k\,.
  \eea
  Again, Eq. \eqref{FlavorGeneral} reduces to $|\nu_\al(\vp,s)\,\rangle \approx \sum_{k}\, \mathcal{U}^\ast_{\al k}\,|\nu_k(\vp,s)\,\rangle$ which is consistent with Eq. \eqref{FlavorState}. If the neutrinos are ultra-relativistic, the condition $|m_k^2-m_{\nu_\al}^2| \ll E_k^2$ is automatically satisfied. On the other hand, as far as the condition $|m_k^2-m_{\nu_\al}^2| \ll E_k^2$ is satisfied, it follows that Eq. \eqref{FlavorGeneral} will certainly reduce to Eq. \eqref{FlavorState} even if neutrinos are non-relativistic.

\end{itemize}

We thus see that neutrino flavor states are approximately well-defined only if neutrinos are ultra-relativistic $(\,m_k,\, m_{\nu_\al} \ll |\vp|\,)$
or the mass differences are negligible compared to energy $(\,|m_k^2-m_{\nu_\al}^2| \ll E_k^2\,)$. These two special cases are slightly different from the two conditions $m_k \ll |\vp|$ and $|m_k^2-m_j^2| \ll |\vp|^2$ derived in \cite{Giunti1992} based on some inconsistent treatments. While both of the two conditions we obtained for approximate neutrino flavor states depend on the indefinite mass $m_{\nu_\al}$,
it is precisely under such conditions that the effect of $m_{\nu_\al}$ becomes negligible. Thus, these two conditions make sense and are consistent with the expectation that the indefinite mass $m_{\nu_\al}$ should be unobservable.

In the above proofs, we have not taken into account of the wave-packet nature of the propagating neutrinos which has been considered in the intermediate (quantum-mechanical) wave-packet model \cite{Kayser,CG1991} or the external (field-theoretical) wave-packet model \cite{CG1993}. Of course, both of the internal and external wave-packet models avoid the neutrino flavor states. But the main purpose of introducing wave-packets was to address the coherence issues in neutrino oscillations and thereby provide a better derivation of oscillation probability relevant for experiments. Even if we consider the wave-packets, it would not help to provide a better construction of neutrino flavor states. So constructing neutrino flavor states and taking into account of the wave-packet nature of neutrinos are two separate issues. For the purpose of the arguments made in this section, it is reasonable to ignore the wave-packets and assume that neutrinos carry a definite momentum $|\vp|$. Furthermore, we remark that while our proofs focused on the three active neutrinos, the two conditions for approximate neutrino flavor states remain valid even if there are sterile neutrinos.

In reality, both of the terrestrial and cosmic neutrinos (except the cosmic neutrino background) are ultra-relativistic. Thus, it is safe enough
for one to compute the transition probability $P_{\al\beta}= |\langle \nu_\beta|\nu_\al (t)\rangle|^2$ through the neutrino flavor states in Eq. \eqref{FlavorState}. This is true for both light active and sterile neutrinos.

Recently, in order to accommodate the anomalies from LSND \cite{LSND} and MiniBooNE \cite{MiniBooNE}, a 3+1+1 scenario has been proposed. In this scenario, the three light active neutrinos mix with one light (eV) and one heavy (up to GeV) sterile neutrinos \cite{NelsonZurek}. Given that the mean neutrino energy in LSND is only 40 MeV, it is obvious that the heaviest mass eigenstate is either non-relativistic or not excited. But according to the proof above, the neutrino flavor state $|\nu_\al \rangle$ is not well-defined for a non-relativistic neutrino. So it is unclear if one can still calculate the transition probability $P_{\al\beta}= |\langle \nu_\beta|\nu_\al (t) \rangle|^2$ through the neutrino flavor states in Eq. \eqref{FlavorState}.


\section{A New Interpretation of Neutrino Flavor States}

In the Standard Model (SM), the charged and neutral current interactions of neutrinos are given by
\bea
\label{LCC}
\mathcal{L}_{\textrm{CC}} &=& \frac{g}{\sqrt2} \left(\,
\overline{\nu}_\al \, \gamma^\mu \, W^+_{\mu}\,L\,l_\al +
\overline{l}_\al\, \gamma^\mu \,  W^-_{\mu}\,L\,\nu_\al
\,\right) \,,\\
\label{LNC}
 \mathcal{L}_{\textrm{NC}} &=& \frac{g}{2 \cos \theta_{_\textrm{W}}} \,\overline{\nu}_\al
\, \gamma^\mu \,  Z_{\mu}\,L\, \nu_\al \,,
\eea
where $\theta_{_\textrm{W}}$ is the Weinberg angle,\;$L=(1-\gamma^5)/2$\, is the left-handed chirality operator, and $l_\al$ are the charged-leptons . From Eq. \eqref{LCC} and Eq. \eqref{LNC}, we commonly say that it is the
flavor neutrinos that are involved in weak interactions, so the concept of neutrino flavor states seems to be physically indispensable.

Nevertheless, we should distinguish the concepts of neutrino flavor fields and flavor states. The flavor fields are dictated by
the fundamental Lagrangian which describes the weak interactions. The flavor states are then generated by acting the creation operators
on the vacuum. These creation operators are derived from the second quantization of the flavor fields. However, due to neutrino mixing,
the flavor fields do not have definite masses. Mathematically, it is consistent to second-quantize a field even if it has an indefinite mass. But it is unclear what is the physical relevance of such a second quantization. As we have shown in the previous section, the neutrino flavor states constructed via the second quantization of flavor fields are approximately well-defined only under some special conditions. In general, the second quantization of a field with indefinite mass simply has no physical relevance. As a result, we suggest to deny the physical significance of neutrino flavor states which are generated from the second quantization of the flavor fields.

One may ask: If we deny the physical significance of neutrino flavor states, how do we describe weak interactions that involve flavor neutrinos?
We can adopt the following prescription. Since the relation in Eq. \eqref{unitary} diagonalizes the free Hamiltonian of neutrinos, we can actually insert this relation into Eq \eqref{LCC} and Eq. \eqref{LNC} so that they becomes:
\bea
\label{LCC_Mass}
\mathcal{L}_{\textrm{CC}} &=& \frac{g}{\sqrt2} \left\{\,
   \left(\,\sum_{j}\, \mathcal{U}^{\ast}_{\al j}\, \overline{\nu}_j\,\right)  \,\gamma^\mu \, W^+_{\mu}\,L\,l_\al +
\overline{l}_\al\, \gamma^\mu \,  W^-_{\mu}\,L\, \left(\,\sum_{k}\, \mathcal{U}_{\al k}\, \nu_k\,\right)\,\right\} \,,\\
\label{LNC_Mass}
 \mathcal{L}_{\textrm{NC}} &=& \frac{g}{2 \cos \theta_{_\textrm{W}}} \, \left(\,\sum_{j}\, \mathcal{U}^{\ast}_{\al j}\, \overline{\nu}_j\,\right)
\, \gamma^\mu \,  Z_{\mu}\,L\, \left(\,\sum_{k}\, \mathcal{U}_{\al k}\, \nu_k\,\right) \,.
\eea
What we have just done was simply a unitary transformation, and so the physics in Eq. \eqref{LCC_Mass} and Eq. \eqref{LNC_Mass} is exactly the same as that in Eq. \eqref{LCC} and Eq. \eqref{LNC}. But the physical interpretation is changed.

Now, the Lagrangian for weak interactions involves only neutrino mass fields. For a given weak process, a specific linear combination of the mass fields is involved, which is dictated by Eq. \eqref{LCC_Mass} and Eq. \eqref{LNC_Mass}. Since the neutrino mass fields have definite masses, it is physically relevant to second-quantize them and construct the Fock space of neutrino mass eigenstates from the physical vacuum $|0\rangle$. Accordingly, when we compute any cross-section of weak interactions involving neutrinos, only the mass eigenstates will be involved.

For instance, instead of saying that $|W^{+}\rangle$ decays into a positron $|e^+ \rangle$ and an electron-neutrino  $|\nu_e \rangle$,
we will say that $|W^{+}\rangle$ simultaneously decays into the individual final states $|e^+,\,\nu_1 \rangle$, $|e^+,\,\nu_2 \rangle$ and $|e^+,\,\nu_3 \rangle$ with the couplings $\frac{g}{\sqrt2}\,U^{\ast}_{e1}$, $\frac{g}{\sqrt2}\,U^{\ast}_{e2}$ and $\frac{g}{\sqrt2}\,U^{\ast}_{e3}$ respectively. There is a kinematic entanglement between each $|\nu_k \rangle$ component and the positron interacting with it.
So the individual $|\nu_k \rangle$ components typically have different momenta and the differences between their momenta depend on the process in consideration. In order to obtain the correct decay width for $W^{+}$, one would need to sum over all the three squared amplitudes (namely, summing over the neutrino mass field index $k$). Since the mass eigenstates are orthogonal to each other, they do not interfere when we compute the cross-section. Perhaps an economical interpretation is that $|W^{+}\rangle$ decays into a superposition state
$\sum_{k}\, \mathcal{U}^\ast_{e k}\, |\,e^+(p_{e_k}), \,\nu_k (p_{\nu_k})\, \rangle$ with the coupling $\frac{g}{\sqrt2}$,
where $p_{\nu_{k}}$ and $p_{e_k}$ are the momenta of the $k^{\textrm{th}}$ neutrino mass eigenstate and the positron entangled
with it respectively.\footnote{We remark that it only makes sense to talk about the neutrino as a coherent
superposition of mass eigenstates if one makes the (reasonable) assumption
that the neutrino is detected with a sufficiently large momentum uncertainty,
so that the three mass eigenstates can interfere in spite of their different
momenta.} In this example, the specific linear combination of mass eigenstates produced in weak interaction is
$\sum_{k}\, \mathcal{U}_{e k}^\ast\, |\nu_k \rangle$. Similar interpretations can be made for any other weak processes. The corresponding linear combinations of mass eigenstates as well as the couplings involved are all dictated by Eq. \eqref{LCC_Mass} and Eq. \eqref{LNC_Mass}. But in fact, we would only encounter $\sum_{k}\, \mathcal{U}_{\al k}^\ast\, |\nu_k \rangle$ or $\sum_{k}\, \mathcal{U}_{\al k}\, |\overline{\nu}_k \rangle$.

As we just saw, a consistent description of weak interaction is guaranteed even if we completely avoid the concept of neutrino flavors states.
We can always work with the mass eigenstates and bypass all the subtleties arise from the construction of the flavor states.
The only drawback is probably that it is notationally tedious to keep track of all the specific linear combinations of the mass eigenstates.
So one may wonder why don't we use illuminating notations to denote them. And undoubtedly, the following definition with the notation $|\nu_\al \rangle$ is
a useful one:
\bea
\label{FlavorStateDefine}
|\nu_\al \rangle \equiv \sum_{k}\, \mathcal{U}_{\al k}^\ast\, |\nu_k \rangle \,,
\eea
which is precisely Eq. \eqref{FlavorState}.

The purpose of going through the above discussions is to point out that by adopting the prescription of
Eq. \eqref{LCC_Mass} and Eq. \eqref{LNC_Mass}, it is always consistent to write down Eq. \eqref{FlavorState}
and calculate the transition probability. The terminologies such as
electron-neutrino, muon-neutrino and tau-neutrino are completely legitimate. However, one needs to keep in mind that
Eq. \eqref{FlavorState} is only a definition. In other words, the flavor state $|\nu_\al\rangle$ is simply a useful notation to denote a specific
linear combination of mass eigenstates involved in weak interactions. It is not derived from the second quantization of the corresponding neutrino flavor field which has an indefinite mass. According to this new interpretation of neutrino flavor states, they have no physical significance. So there is no need to construct the Fock space of neutrino flavor states from the first principles of quantum field theory.

An immediate implication of this new interpretation is that since Eq. \eqref{FlavorState} is a definition which is always true, it is still
reasonable to calculate the transition probability $P_{\al\beta}= |\langle \nu_\beta|\nu_\al (t) \rangle|^2$ in the 3+1+1 scenario \cite{NelsonZurek}
even though one of the sterile neutrinos maybe non-relativistic. Of course, if the heaviest sterile neutrino is really non-relativistic, then there may be some other complications, such as non-unitarity, when we evaluate $|\langle \nu_\beta|\nu_\al (t) \rangle|^2$ \cite{NelsonZurek}.

\section{Discussions and Conclusions}

In this article, we first gave a detailed review on the current status of the understanding about the neutrino flavor states.
At the end of the review, we were led to conclude that it is currently unclear how to construct a consistent and physically relevant Fock space of neutrino flavor states. We proceeded to prove that if one insists on second-quantizing the neutrino flavor fields and thereby constructing the flavor states, then they are approximately well-defined only when neutrinos are ultra-relativistic or the mass differences are negligible compared to energy.

However, we showed that one can consistently describe weak interactions by only neutrino mass eigenstates.
At the same time, we argued that the second quantization of neutrino flavor fields generally lacks physical relevance because their masses are indefinite. Thus, neutrino flavor states lose their physical significance and they should simply be interpreted as definitions to denote specific linear combinations of mass eigenstates involved in weak interactions. Under this interpretation, there is no physical motivation to construct the Fock space of neutrino flavor states from the first principles of quantum field theory.

It is well-known that the external wave-packet model \cite{CG1993} also denies the physical significance of the neutrino flavor states. In fact, it abandons the flavor states completely. To some extend, our interpretation that neutrino flavor states are only definitions shares the same spirit of the external wave-packet model. One difference (apart from the wave-packet nature) is probably that we admit that the expression $|\nu_\al \rangle = \sum_{k}\, \mathcal{U}_{\al k}^\ast\, |\nu_k \rangle$ is always consistent, while the proponents of the external wave-packet model constructed the model to avoid it.

\begin{acknowledgments}
We would like to thank D. Boyanovsky and J. Kopp for useful discussions.
We also thank the anonymous referee for illuminating comments.
This work was supported in part by the Department of Energy (DE-FG05-85ER40226).
\end{acknowledgments}

\end{document}